\documentstyle[11pt,IAUS212,twoside]{article}

\def\edcomment#1{\iffalse\marginpar{\raggedright\sl#1\/}\else\relax\fi}
\reversemarginpar

\begin{document}
\vspace*{1cm}
\title{Grain Formation in Post-Shocked Wind Collisions of Massive
Binary Systems}

\author{D. Falceta-Gon\c calves, V. Jatenco-Pereira \& Z. Abraham}
\affil{Instituto de Astronomia, Geof\'\i sica e Ci\^encias Atmosf\'ericas - 
USP \\
 Rua do Mat\~ao, 1226 - 05508-090 - S\~ao Paulo - Brazil}


\begin{abstract}
Massive binary star systems are not uncommon, and neither the 
supersonic collision of their winds. In the present work we study 
these shocks and the further consequences on wind structure.
The post-shock gas is a warm and high-density environment, which 
allows dust to form and grow. We show that this growth is fast, of 
just a few hours. An application for $\eta$Car shows that, probably, the
decline of X-rays fluxes observed in its light curve is the consequence
of its high absorption in periodic dust formation events, on the periastron
passage.
\end{abstract}

\section{Introduction}

Many massive binary systems are known, and almost all present 
high-energy fluxes, not originated in the stars, but around 
them, indicating the presence of colliding winds. The source 
of observed X-rays high fluxes should 
be a hot and dense gas, generated by shocks around these objects 
(Zhekov \& Skinner 2000). 
At high temperature and density, free-free 
emission in X-rays becomes highly important (Usov 1992; 
Ishibashi et al. 1999). 
X-ray emissions associated with interacting winds were observed 
on many objects (Corcoran et al. 2001; Thaller 
et al. 2001). Grains also may be formed and grow on post-shocked 
gases (Monnier, Tuthill, \& Danchi 2001) when the heated gas loses 
high amount of energy, mainly by {\it bremsstrahlung}, cools and 
becomes denser. Recent observations in IR supply this idea, showing 
episodic dust formation events in these regions (Marchenko, 
Moffat, \& Grosdidier 1999; Harries, Babler, \& Fox 2000). Some 
binary systems present, periodically, 
a sudden decrease of X-ray emission (Ishibashi et al. 1999). 
A fast dust formation and growth event in these regions due to the 
shocked winds might provoke a considerable increase in extinction 
explaining these features of the light curves. This could cause the 
temporary reduction on X-rays and UV fluxes, which may increase 
{\it a posteriori} with the evaporation of the particles, or 
even by the expansion of the dust shell.

\section{Model and Results: an application to $\eta$Carinae}

The model assumes the colliding winds of two massive stars A and B, 
which have mass loss rates $\dot M_A$ and $\dot M_B$, and wind velocities 
$u_A$ and $u_B$, respectively, orbiting each other with period $T$, 
eccentricity $e$ and distance $D$ at {\it periastron}.
We divide the 
shock evolution in four parts: 
i) the instantaneous collision, neglecting radiative losses, ii) the 
gas cooling at post-shock, iii) the formation of dust and its growth 
and iv) the evolution of the dust shell, and its consequence to the 
light curve observed. The basic physics of the shock is very well known, 
and described by the hydrodynamics equations of: mass continuity, fluid 
momentum and energy. The growth and evaporation rates of the dust particles 
are given by the classic nucleation theory.

$\eta$Car is a super massive star, supposed to be a LBV class star, 
and also is supposed to have a companion, 
probably another massive WR 
or OB star (Damineli et al. 2000). The system total mass, inferred by 
luminosity is $\sim 120 M_\odot$. $\eta Car_A$ is responsible 
for major of this mass and $\eta Car_B$ should have typical WR 
stellar winds, $\dot M \sim  10^{-5} \; M_\odot \; yr^{-1}$ and 
$u_\infty \sim 3\times10^3 \; 
km \; s^{-1}$. Observations indicate that $\eta Car_A$ should 
have $\dot M \sim 10^{-3} M_\odot \; yr^{-1}$ and $u_\infty \sim 500 \; 
km \; s^{-1}$ 
(Hillier 2001; Damineli et al. 2000). For the above parameters and
$\dot M = 5\times 10^{-4} \; M_\odot \; yr^{-1}$ and
$e \sim 0.8$, it was possible to reproduce its observed light curve.

\section{Conclusions}

In this work we present a model for colliding winds where we determined the 
changes in physical parameters, like density, temperature and pressure in the 
post-shock. This gas, cooled and denser after shock, can be the site for 
grains formation. The model was applied to $\eta Car$ which could 
explain: i-) the rapid decline on 
fluxes observed periodically; and ii-) the slightly decrease in opacity
considering the expansion of the dust with the wind. This model agrees 
better with observations for $e \sim 0.8$, and a 
mass loss rate of the primary star 
of $\sim (3 - 5)\times10^{-4} \; M_\odot \; yr^{-1}$.

\acknowledgements
$\; \;$ The authors thank the Brazilian agencies CNPq, \- \- CAPES and FAPESP for
support and also the project PRONEX (41.96.0908.00).

\end{document}